\documentclass{JHEP3}
\let\ifpdf\relax

\usepackage[parfill]{parskip}    
\usepackage{graphicx}
\usepackage{fancyvrb}
\usepackage{makeidx}
\usepackage{latexsym,amssymb,amsmath,graphicx, makeidx}
\usepackage{ifpdf}

\newcommand{\be}{\begin{equation}}
\DefineVerbatimEnvironment{code}{Verbatim}{fontsize=\small}
\newcommand{\ee}{\end{equation}}
\newcommand{\bi}{\begin{itemize}}
\newcommand{\ei}{\end{itemize}}
\makeindex

\newcommand{\sigSI}{\sigma_{\rm SI}}

\newcommand{\sigSD}{\sigma_{\rm SD}}
\newcommand{\lsim}{{\ensuremath \raisebox{2pt}{$<$}}_{\!\!\!\!\!\sim}}

\newcommand{\etmiss}{\not{\!\!E}_{T}}

\newcommand{\zp}{Z^{\prime}}

\newcommand{\gzp}{g_{Z'}}
\newcommand{\gzpa}{g_{Z' 5}}
\newcommand{\mzp}{M_{Z'}}
\newcommand{\mchi}{M_{\chi}}
\newcommand{\gd}{g_{D}}
\newcommand{\gda}{g_{D 5}}

\preprint{PI-PARTPHYS-258, UMD-DOE/ER/40762-514}

\title{Light dark matter and $Z'$ dark force at colliders}

\author{Haipeng An$^{a,b}$, Xiangdong Ji$^{a,c,d}$, and Lian-Tao Wang$^{e,f}$\\
        $^a$Maryland Center for Fundamental Physics and Department of Physics, University of
Maryland, College Park, Maryland 20742, USA\\
        $^b$Perimeter Institute, Waterloo, Ontario N2L 2Y5, Canada\\
        $^c$Shanghai Laboratory for Particle Physics and Cosmology, and Department of Physics, Shanghai Jiao Tong University, Shanghai 200240, China \\
        $^d$Center for High-Energy Physics and Institute of Theoretical Physics, Peking
University, Beijing 100871, China  \\	
        $^e$Enrico Fermi Institute and Department of Physics, University of Chicago, Chicago, IL, 60637\\
       $^f$Kavli Institute for Cosmological Physics, University of Chicago, Chicago, IL, 60637\\
       E-mail: \email{han@perimeterinstitute.ca, xji@umd.edu, liantaow@uchicago.edu}
       }

\abstract{
Light Dark Matter, $<10$ GeV, with sizable direct detection rate is an interesting and less explored scenario. Collider searches can be very powerful, such as through the channel in which a pair of dark matter particle are produced in association with a jet.  It is a generic possibility that the mediator of the interaction between DM and the nucleus will also be accessible at the Tevatron and the LHC. Therefore, collider search of the mediator can provide a more comprehensive probe of the dark matter and its interactions. In this article, to demonstrate the complementarity of these two approaches,  we focus on the possibility of the mediator being a new $U(1)'$ gauge boson,  which is probably the simplest model which allows a large direct detection cross section for a light dark matter candidate. We combine searches in the monojet+MET channel and dijet resonance search for the mediator. We find that for the mass of $Z'$ between 250 GeV and 4 TeV, resonance searches at the colliders provide stronger constraints on this model than the monojet+MET searches.
}

\begin{document}
\maketitle
\tableofcontents

\section{Introduction}
\label{sec: introduction}

The identity of the Cold Dark Matter (CDM) in the universe is one of the outstanding mysterious.
Many possible models of dark matter have been proposed.
In many promising scenarios, such as those realizing the so-called WIMP miracle, dark matter posses non-vanishing couplings to the Standard Model (SM) particles. Experiments designed to observe dark matter through these interactions are crucial in probing the properties of the dark matter.

One of such promising approaches is the direct detection,  with many results being reported recently~\cite{Bernabei:2008yi,Ahmed:2010wy,Akerib:2010pv,Aprile:2011hi,Aalseth:2011wp,Angloher:2011uu}. The current direct detection of dark matter has roughly two frontiers. The first frontier is for dark matter mass $M_\chi$ in the range of $100$s GeV $-$ TeV. As a frequently highlighted result,   the constraint on the spin-independent dark matter nucleon cross section, $\sigSI$,  is around $10^{-44}-10^{-45}$ cm$^{2}$. Along this frontier, the direct searches are starting to set interesting constraints on the "conventional" WIMP candidate, such as the LSP dark matter of supersymmetry \cite{Goldberg:1983nd,Ellis:1983ew} or KK dark matter of models with extradimension(s) \cite{Cheng:2002ej,Servant:2002aq}. The second frontier is in the regime of light, $M_\chi \leq 10$ GeV, dark matter. In this regime, the momentum of the dark matter in the galactic halo is relative small, leading to suppressed recoil energy of the detector nuclei. Therefore, the bound on the dark matter nucleon cross section is much weaker, dropping from $ \sigSI \lsim 10^{-42}$ cm$^2$ for $M_\chi \simeq 10$ GeV to $ \sigSI \lsim 10^{-39}$ cm$^2$ for $M_\chi \simeq 6$ GeV, for example. For the same reason, there is also a similar feature in the reach for spin-dependent interactions, $\sigSD$. The possibility of light dark matter within this mass regime has received considerable attention recently \cite{Feng:2008ya,Petriello:2008jj,Feng:2008dz,Zurek:2008qg,Kaplan:2009ag,Feldman:2010ke,Fitzpatrick:2010em,Kuflik:2010ah,Andreas:2010dz,Graham:2010ca,Chang:2010yk,Essig:2010ye,An:2010kc,Cohen:2010kn,Das:2010ww,Hooper:2010uy,Fitzpatrick:2010br,Belanger:2010cd,Foot:2010hu,Kang:2010mh,Barger:2010mc,Belikov:2010yi,Gunion:2010dy,Draper:2010ew,Vasquez:2010ru,Fornengo:2010mk,Schwetz:2010gv,Dutta:2010va,Feng:2011vu,Cai:2011kb}.

High energy collider experiments are crucial in the search of dark matter, independent of astrophysical uncertainties.  Dark matter particles can be pair produced at colliders, and they will escape without interacting with the detector. Therefore, the basic signature of dark matter is jet(s) (or $\gamma$)$+ \etmiss $. If dark matter is light, the production rate is determined dominantly by the cut on ${\etmiss}$ and relatively independent of $M_\chi$. Therefore, Collider searches can be much more powerful and provide crucial complementary information in the case of light dark matter where direct detection has its intrinsic limitations.

Some model dependence in making the connection between direct detections and collider searches is unavoidable. The momentum transfer involved in the direct detection experiments is tiny, $\lsim 10$(s) MeV. Therefore, we can typically integrate out the particle which mediates the interaction between DM and SM particles. The resulting operators, of the form $J_{\rm SM} \cdot J_\chi $, provide adequate description for direct detection. The simplest approach is to use the same operator in collider studies, treating them as contact interactions \cite{Beltran:2010ww,Goodman:2010yf,Bai:2010hh,Goodman:2010ku,Goodman:2010qn,Fortin:2011hv,Rajaraman:2011wf,Shoemaker:2011vi}. At the same time, the mediator is not necessary heavy and therefore not in the decoupling limit. In this case, it is not a very good approximation to integrate out the mediator at collider energies. Such effects have already been pointed out and studied  \cite{Beltran:2010ww,Goodman:2010yf,Bai:2010hh, Goodman:2010ku,Goodman:2010qn,Fortin:2011hv,Rajaraman:2011wf,Shoemaker:2011vi}. Moreover, since the mediator is within the reach of collider energies, there are additional channels we can use to search for it, such as resonant production followed by the decay back to SM states.

The main goal of this paper is to extend the earlier studies  \cite{Beltran:2010ww,Goodman:2010yf,Bai:2010hh, Goodman:2010ku,Goodman:2010qn,Fortin:2011hv,Rajaraman:2011wf,Shoemaker:2011vi} by including the effect and signature of a mediator which is non-decoupling. In particular, we combine the mono-jet + MET search with direct collider searches for the mediator, and demonstrate that they are both powerful and complementary.  In the light of the strength of collider searches in the small $M_\chi$ and large $\sigSI$ frontier, we concentrate on mediator with proper coupling to both SM and DM which give rise to such large cross sections. The most obvious example is the scenario in which dark matter is a Dirac fermion, and its interaction with the nucleus is spin-independent and is through a spin-1 ($Z'$) mediator. There is a corresponding example with  a Majorana fermion dark matter  and  a $Z'$ mediator, in which the dominant dark matter nucleus interaction is spin dependent.

The rest of the paper is organized as follows. In Sec.~\ref{Sec:GenericZp}, we discuss the generic properties of $Z'$ model with all possible renormalizable couplings to quarks and dark matter.We also illustrate the transition from on-shell production of $Z'$ to contact interactions. We emphasize that in order to simulate the monojet and dijet constraint correctly with a relatively broad resonance, one has to use kinetic width, which is discussed in detial in Appendix A. In Sec.~\ref{Sec:leptophobic}, we discuss constraints and reaches from Tevatron and LHC in search for excess in the jet$+\ \etmiss$ channel. We  translate the results to constraints on direct detection cross sections of elastic collision between nucleon and dark matter. Results for both $\sigSI$ and $\sigSD$  are presented.
In Sec.~\ref{Sec:dijet}, we discuss the dijet constraint on $Z'$ mass and its couplings to the quarks. We find that if the couplings between DM and $Z'$ is comparable or smaller than the couplings between quarks and $Z'$,  the constraint from dijet is stronger than from monojet.
The constraints on couplings of the new gauge boson are shown in Appendix B.

\section{Leptophobic $Z'$ model, direct detection, and searches at colliders}
\label{Sec:GenericZp}

If the interactions between dark matter and SM quarks are mediated by $Z'$, including all renormalzable interactions, the Lagrangian can be written as
\begin{eqnarray}\label{lgrg1}
{\cal L} &=& Z'_{\mu} [(g_{Z'} \bar q \gamma^{\mu} q + g_{Z'5} \bar q \gamma^{\mu}\gamma_{5} q)
+ ( g_{D} \bar \chi \gamma^{\mu} \chi + g_{D5} \bar \chi \gamma^{\mu}\gamma_{5} \chi )]\ ,
\end{eqnarray}
where $q$ and $\chi$ are denoting SM quarks and dark matter particles, respectively.
In general, $Z'$ can also couple to leptons, and we can parameterize such extensions with $B-xL$ type of couplings. In this case, the search of dilepton resonance provides the most obvious probe and its reach depends sensitively on $x$, which is not related to the direct detection of dark matter.  In this paper, we will concentrate on the minimal case of a leptophobic $Z'$.

In direct detection experiment, the momentum transfer between dark matter particle and the nucleus can be estimated as $M_N M_\chi v \cos\theta/(M_N+ M_\chi )$. Therefore, we can see that the momentum transfer is limited by the mass of nucleus and the speed of dark matter and cannot be larger than around 0.1 GeV. Therefore, in the case that $M_{Z'}$ is larger than a few GeV, the interaction between nucleus and dark matter can be described by a contact interaction which depends only on the Wilson coefficient $g_{Z'}g_D/M_{Z'}^2$.
\begin{table}[h!]
\begin{center}
\begin{tabular}{lllc}
~&~ Operator                                                            ~&~ Structure ~&~ DM-nucleon Cross Section\\
\hline
\rule{0cm }{0.7cm} $O_1$  \vspace{0.1cm}
 ~&~ $\bar q\gamma^\mu q \bar\chi\gamma_\mu\chi$                             ~&~ SI, MI ~&~  $\frac{9 \gzp^2 \gd^2 M_N^2 M_\chi^2}{\pi \mzp^4 (M_N+M_\chi)^2}$     \\
\hline
\rule{0cm }{0.7cm} $O_2$  \vspace{0.1cm}
 ~&~ $\bar q\gamma^\mu q\bar\chi\gamma_\mu\gamma_5\chi$                      ~&~ SI, MD ~&~ $\sim v^2$ \\
\hline
\rule{0cm }{0.7cm} $O_3$  \vspace{0.1cm}
~&~ $\bar q\gamma^\mu\gamma_5 q \bar\chi\gamma_\mu\chi$                      ~&~ SD, MD ~&~ $\sim v^2$\\
\hline
\rule{0cm }{0.7cm} $O_4$  \vspace{0.1cm}
~&~ $\bar q\gamma^\mu\gamma_5 q\bar\chi\gamma_\mu\gamma_5\chi$              ~&~ SD, MI  ~&~ $\frac{3 \gzpa^2 \gda^2 (\Delta\Sigma)^{2} M_N^2M_\chi^2}{\pi \mzp^4 (M_N+M_\chi)^2}$
\end{tabular}
\end{center}
\caption{\label{tab:zprime_int} Effective operators for dark matter direct detection after integrating out the $Z'$, and the corresponding DM nucleon cross sections.  We also show whether the dark matter nucleus cross section is independent (SI) or dependent (SD) on the nucleus spin, and whether it is dependent on the dark matter momentum (MD) or not (MI). $M_N$ denotes the nucleon mass. $\Delta\Sigma$ is defined as $\langle N|\sum_{q}\bar q \gamma_{\mu}\gamma_{5} q |N\rangle = \Delta\Sigma \bar U_{N}\gamma_{\mu}\gamma_{5} U_{N}$, where $U_{N}$ is the wave function of the nucleon. $O_3$ and $O_4$ are parity-odd operators inducing velocity dependent interaction between DM and nucleons. $v$ is the velocity of DM in the lab frame.}
\end{table}

We begin with a review of the connection between direct detection and various possible $Z'$ interactions. After integrating out the $Z'$, we obtain 4 different effective operators. We collect them, as well as their corresponding DM-nuleon cross section and the relevant properties of DM-nucleus cross section, in Table~\ref{tab:zprime_int}.  It is well known that dark matter direct detection rate is enhanced by coherent scattering off the nucleus if the dark matter nucleon interaction is spin independent. Typical velocity of the dark matter in the galactic halo is $v \sim 10^{-3}$. Any process depending on the momentum exchange $|\Delta \vec{p}_N| = |\Delta \vec{p}_\chi | \sim |\vec{p}_\chi |$ is suppressed by an additional factor of $v^2$.

In the case that $Z'$ couples only to vector currents of both quark and dark matter fields ($O_1$), the cross section is both spin independent, and it is unsuppressed by momentum exchange.  Using some typical values of gauge couplings and $\mzp$, we can estimate
\begin{equation}\label{WIMPcx}
\sigma_{\rm SI} = \frac{9g_{Z'}^{2}g_{D}^{2}M_{N}^{2}M_{D}^{2}}{\pi M_{Z'}^{4}(M_{N}+M_{D})^{2}} \simeq 3.9 \times 10^{-39} \text{cm}^{2} \left(\frac{g_{Z'}}{0.3} \right)^2  \left(\frac{g_{D}}{0.3} \right)^2 \left(\frac{200 \text{\ GeV}}{\mzp} \right)^4.
\end{equation}
Therefore, this is a very plausible way of realizing the large cross section scenario we are interested in probing. We are going to to mainly focus on this case in this paper.

It is possible that $O_1$ is absent or strongly suppressed. An important example is when the dark matter particle is a Majorana fermion. In this case, dark matter can only form axial vector current and $O_2$ and/or $O_4$ describes the direct detection, with $O_4$ typically being the dominant one since $O_2$ is further suppressed by momentum exchange. We will consider this case separately in Sec. \ref{sec:sd}.

In most of the cases considered in this paper, the $U(1)'$ is anomalous, and  spectator fermions need to be introduced to cancel the anomaly. In this case, the upper bound for the mass of spectating fermion is~\cite{Preskill:1990fr}
$M_{\rm spectator} < (64\pi^{2}/g_{Z'}^{3})M_{Z'}\ $,
where $g_{Z'}$ is the coupling between $Z'$ and SM fermions and $M_{Z'}$ is the mass of $Z'$. Potential signals from these states provide further information about this scenario. To focus on the most generic phenomenology, we will not include explicit spectators in this work.

The size of the signal of dark matter depends on  sizes of couplings $\gd$ and $\gzp$. In principle, they don't have to be related. However, if DM particle is part of the spectator, for example in the scenario of Ref.~\cite{Buckley:2011mm}, the couplings will have to be of the same order. On the other hand, for example, they might be introduced through effective interactions~\cite{Fox:2011qd} and they can be quite different.
In this work, to illustrate the basic features of our results, we first  assume $g_{Z'}=g_{D}$. From the various collider processes we get upper bounds on $g_{Z'}$ ($g_{D}$), with which we then set the upper bound on direct detection cross section. We then proceed to investigate how the constraints change if we relax this assumption and allow $\gzp/\gd$ to vary.

In collider production of dark matter particles, the relevant momentum scale, $q$, can be comparable
with $\mzp$ and the dependence of the signal rate on the model parameters are in general more complicated, thus
the contact interaction approximation is not always a good one.  To demonstrate this, we study the production
cross section as a function of $Z'$ model parameters, particularly the $Z'$ mass, fixing the combination
 $g_{Z'}g_D/M_{Z'}^2 = 1/\Lambda^2$ and hence the direct detection rate. We further assume $g_{Z'}=g_D$ for simplicity.
We have at parton level,
\begin{equation}\label{propagator}
\hat\sigma_{\rm collider} \propto \frac{g_{Z'}^4}{(q^2 - M_{Z'}^2)^2 + q^2 \Gamma_{Z'}^2(q^2)} = \frac{1}{(q^2 \frac{\Lambda^2}{\mzp^2} - \Lambda^2)^2 + a^2\frac{(q^2)^2}{(12\pi)^2}} \ ,
\end{equation}
where $\Gamma_{Z'} (q^2)$ is the width of $Z'$ with $M_{Z'}^2$ replaced by $q^2$, and $a$ is a constant depending on couplings and the number of degrees of freedom.

\FIGURE[h!]{
\includegraphics[scale=0.7]{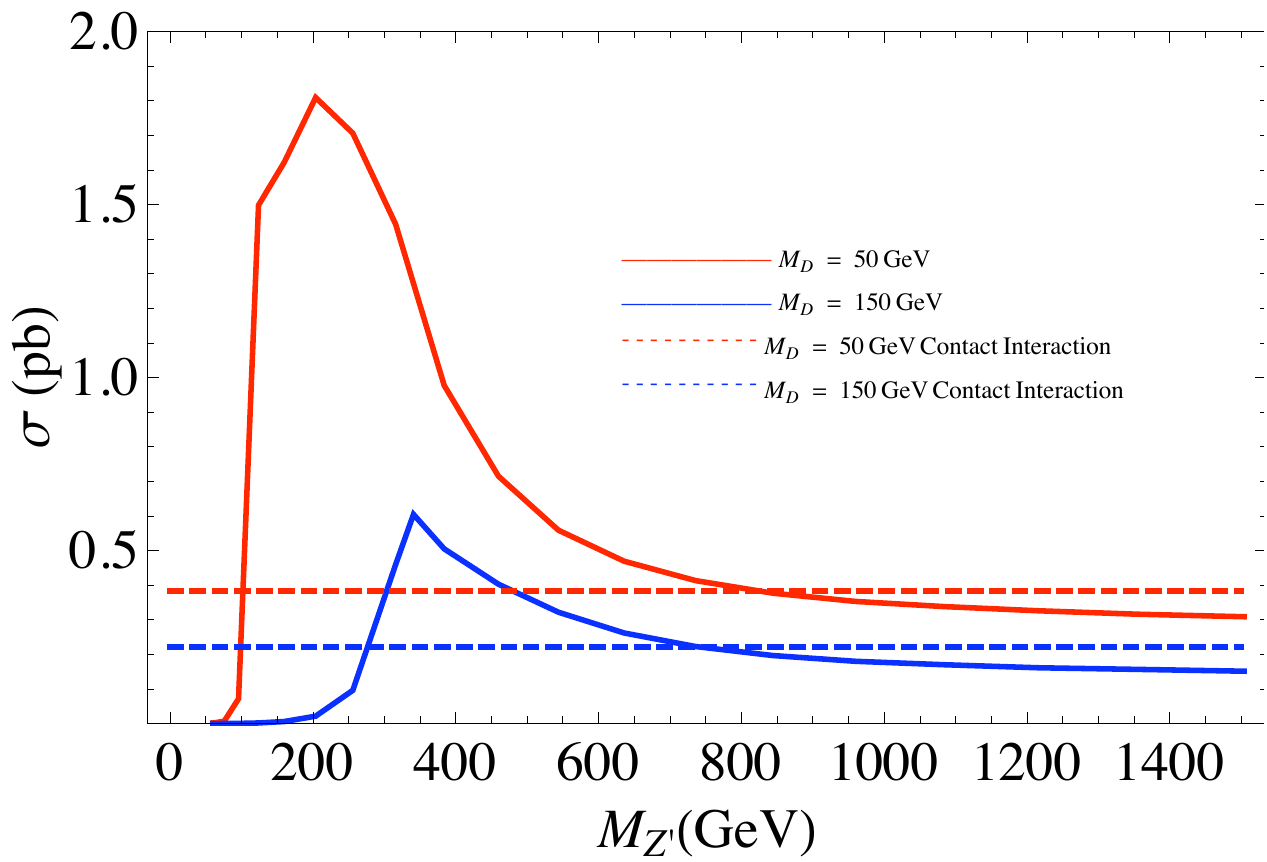}
\caption{ Cross section for monojet + MET at Tevatron  as a function of $M_{Z'}$ with different dark matter masses.   $g_{Z'}/M_{Z'}$, and hence the direct detection cross section $\sigma_{\rm direct}$, is fixed. We assume $g_{Z'}=g_D$.
\label{fig:decoupling}}}

Fig.~\ref{fig:decoupling} shows the monojet + MET cross section (to be discussed in detail in the next section) at Tevatron as a function of $M_{Z'}$ with $\Lambda$ fixed to 300 GeV, for two different values of $\mchi$. For comparison, we have also shown the production rates predicted by using a contact interaction with coefficient $1/\Lambda^2$. When $M_{Z'}<2 \mchi$, $Z'$ cannot be on shell in this process. The parton level cross section for $ q \bar{q} \to (Z^{\prime *} \to \chi \chi) +X $ can be written as
\begin{equation}
\hat\sigma_{\rm collider} \sim \frac{1}{\left( \frac{4\mchi ^2}{M_{Z'}^2} \right)^2 \Lambda^4+a^2\frac{(4 \mchi^2)}{(12\pi)^2}}\ .
\end{equation}
The cross section is suppressed by $(M_{Z'}^2/4 \mchi^2)^2$ at small $\mzp$.
In the case that $M_{Z'}>2 \mchi $, $Z'$ can be produced on shell. Using the narrow width approximation,
\begin{equation}
\hat\sigma \sim \hat\sigma_{\rm prod}(Z')\times {\rm Br}(Z'\rightarrow \chi\bar\chi)\ .
\end{equation}
Therefore, the threshold behavior near $\mzp \sim 2 \mchi$ in Fig.~\ref{fig:decoupling} is due to the enhancement from resonant production.

For larger $\mzp$, the interaction between quarks and dark matter becomes a contact interaction which is shown by the flat tail in Fig.~\ref{fig:decoupling}. For very large $\mzp$, typical $q^2$ is independent of $\mzp$ and constrained by parton density to be $q^2 \ll \mzp$. The production rate asymptotes to a constant, which is less than the prediction from contact interaction due to the width term on the denominator in Eq.~(\ref{propagator}).

In Eq.~(\ref{propagator}), the kinetic width of the intermediate particle is used instead of the Breit-Wigner approximation. In our discussions, since $Z'$ is assumed to couple to all flavors of quarks universally and the gauge coupling is allowed to vary in a wide range, the width of $Z'$ can be broad.  Therefore, the Breit-Wigner approximation with a large constant width cannot be seen as a good approximation in this case. A detailed discussion of this effect can be found in Appendix~\ref{Appendix:A}.

\section{Monojet + MET searches at the Tevatron and the LHC}
\label{Sec:leptophobic}

We consider dark matter particle production in association with a jet radiated from the initial state parton.
Some examples of the Feynman diagrams are shown in Fig.~\ref{fig:monojet-diagram}. This process, arguably being the most model
independent, has been studied \cite{Beltran:2010ww,Goodman:2010yf,Bai:2010hh,Goodman:2010ku,Goodman:2010qn,Fortin:2011hv,Rajaraman:2011wf,Shoemaker:2011vi,CDF7ifb:001}.
We present our study within our $Z'$ framework, taking into account the effects of kinematic $Z'$ width as described in
Appendix A. The experimental data is from Tevatron and LHC searches with $L \sim 1$ fb$^{-1}$. The result is shown as
the constraints on direct detection cross section, whereas the constraints on the couplings are presented in Appendix B.

\subsection{Constraints for the spin-independent case}
\label{Sec:SI}

\FIGURE[h!]{
\includegraphics[scale=1]{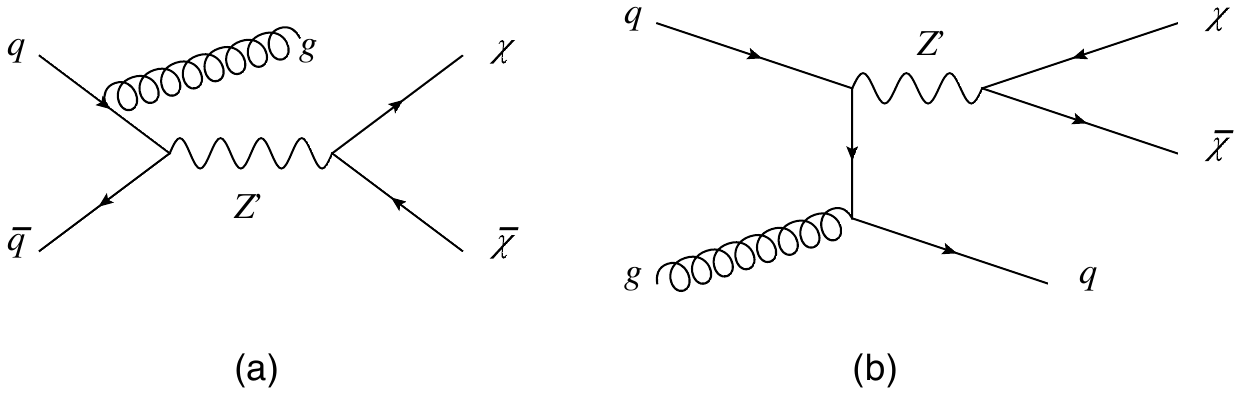}
\caption{ Dominant parton level diagrams for $p\bar p\rightarrow$ monojet + MET.  (b) is also the dominant parton level process for $pp\rightarrow$ monojet + MET in LHC.
\label{fig:monojet-diagram}}}

Most of the existing searches for new physics in the monojet+MET channel are carried out in the  context of the large extra dimension (LED) model~\cite{ArkaniHamed:1998rs}. We will use those results to set limits on our $\zp$ model. Being a non-renormalizable model, the $p_T$ (and $\etmiss$) distribution predicted by LED is somewhat harder than the $\zp$ model, and different cuts can be chosen to improve the sensitivity.

The strategy we use to get constraint from monojet search on the $Z'$ model follows from Ref.~\cite{Fox:2011pm}, by defining a $\chi^2$,
\begin{equation}
\chi^2 = \frac{[N_{\rm obs}-N_{\rm SM}-N_{Z'}]^2}{N_{Z'}+N_{\rm SM}+\sigma_{\rm SM}^2} \ ,
\end{equation}
where $N_{\rm obs}$ is the number of events observed in the experiment, $N_{\rm SM}$ is the prediction
from SM, $\sigma_{\rm SM}$ is the uncertainty of the SM prediction including statistic and systematic uncertainties.
Then we require $\chi^2 < 2.71$ to get 90\% constraint on the contribution from the $Z'$ model, $N_{\rm Z'}$.

The  selection cuts used by various searches, which we also adopt, are listed in  Table.~\ref{ATLAS_cuts}.
CDF group used 1 fb$^{-1}$ of data~\cite{Aaltonen:2008hh}, and
two sets of cut on $\etmiss$ (MET) and the leading jet were used. The constraints on the $\zp$ model from LowPT cut is always stronger than the one
from HighPT cuts. Therefore, we will use the lower PT cut to set limits. With this set of cuts, 8449 events have been observed,
which is consistent with the SM background prediction of 8663$\pm$332.
ATLAS and CMS have published their analysis of monojet + MET with a luminosity of 1 fb$^{-1}$, we will concentrate on the
ATLAS search in our study \cite{Atlasmonojet1fb}. They did analysis with three different PT cuts,
namely LowPT, HighPT and veryHighPT, as shown in Table~\ref{ATLAS_cuts}.

\begin{table}[htdp]
\begin{center}
\begin{tabular}{|c|l|}
\hline
CDF 1 fb$^{-1}$~&~ $\not\!\! E_T > 80 {~\rm GeV}$, $p_T(j_1)>$80 GeV. \\
                             ~&~ Events with additional jets are vetoed if \\
                             ~&~  $p_T(j_2)>30$ GeV or $p_T(j_3)>$  20 GeV. \\
\hline
CDF 6.7 fb$^{-1}$ ~&~ $\not\!\! E_T > 60 {~\rm GeV}$,  $p_T(j_1)>$60 GeV. \\
                            ~&~ A second jet is allowed if 20 GeV $< p_T(j_2)<$ 30 GeV.                   \\
\hline
ATLAS LowPT ~&~ $\not\!\! E_T > 120 {~\rm GeV}$, one jet $p_T(j_1) > 120 {~\rm GeV}$,
                $|\eta(j_1)|<2$. \\
            ~&~    Events with a second jet  are vetoed \\
           ~&~    if $p_T(j_2)>30{~\rm GeV}$ and $|\eta(j_2)|<4.5$. \\
\hline
ATLAS HighPT ~&~ $\not\!\! E_T > 220 {~\rm GeV}$, $p_T(j_1) > 250 {~\rm GeV}$, $|\eta(j_1)|<2$. \\
               ~&~ Event is vetoed if there is  a second jet with $p_T(j_2) > 60 {~\rm GeV}$ \\
               ~&~ or $\Delta\phi(j_2,\not\!\!E_T)<0.5$ and $|\eta(j_2)|<4.5$.\\
               ~&~  Any additional jet with $|\eta(j_3)|<4.5$ must have $p_T(j_3)<30$ GeV. \\
\hline
ATLAS vertHighPT ~&~  $\not\!\! E_T > 300 {~\rm GeV}$, one jet with $p_T(j_1) > 350$ GeV. \\
                     ~&~ $|\eta(j_1)|<2$, and events are vetoed if there is a second jet with  \\
                     ~&~ $|\eta(j_2)|<4.5$ and with either $p_T(j_2)>60$ GeV or $\Delta(j_2,\not\!\! E_T)<0.5$. \\
                     ~&~ Any further jets with $|\eta(j_3)|<4.5$ must have $p_T(j_3)<30$ GeV. \\
\hline
\end{tabular}
\end{center}
\caption{Different cuts defined by CDF and ATLAS in monojet + $\not\!\! E_T$ searches.  $j_1$ and $j_2$ denote leading and second leading jets respectively.  $j_3$ labels any other jets. }
\label{ATLAS_cuts}
\end{table}
In our analysis, we generated events at parton level, and applies selection cuts. A full analysis would also include further steps of parton shower and merging it with hadronization. Such studies have been carried out in the case of contact operator \cite{Beltran:2010ww,Goodman:2010yf,Bai:2010hh,Goodman:2010ku,Goodman:2010qn,Fortin:2011hv,Fox:2011pm}, for which
the results from different couplings can be obtained from simple scaling. In our case, we in principle have to carry out full simulations for signal at every possible mass and coupling in our scan \footnote{We note that there is no simple dependence of the monojet cross section on  $\mzp$ and $\gzp$ which is valid for the full parameter space.  For example, $\gzp$ enters the $Z'$ width. Therefore,  the cross section is not a simple function $\propto \gzp^2$, especially when the $\zp$ is broad.Therefore, we can not obtain correct results just by scaling from the simulation using a particular set of ($\gzp$, $\mzp$).   }. Instead, we match our parton level result in the contact-interaction limit (large $\mzp$) with the full studies in Ref.~\cite{Goodman:2010yf} (Tevatron) and  Ref.~\cite{Fox:2011pm} (LHC) and obtain a scale factor. This scale factor is then applied to our model for general $\mzp$ and $\gzp$. Although just an approximation, this does capture the main difference between a parton level and a full study in our case.
For the CDF simulation, the cut we are using is that at parton level, the $p_T$ of the jet should be larger than 80 GeV. For ATLAS simulation, we require that the $p_T$ of the jet should be larger than 120 GeV, 220 GeV and 300 GeV for LowPT, HighPT and veryHighPT, respectively. For CDF, if the mass of DM is smaller than 100 GeV, the scale factor is about 0.6, about 0.4 for ATLAS LowPT cut, and about 0.5 for HighPT and veryHighPT cuts. The difference between parton level simulation and the full study is mainly caused by the fact that after parton shower, the energy and $p_T$ of the clustered just tend to be smaller than the original parton. 
Hence,  the scale factor is almost always smaller than one.

\FIGURE[th]{
\includegraphics[scale=1]{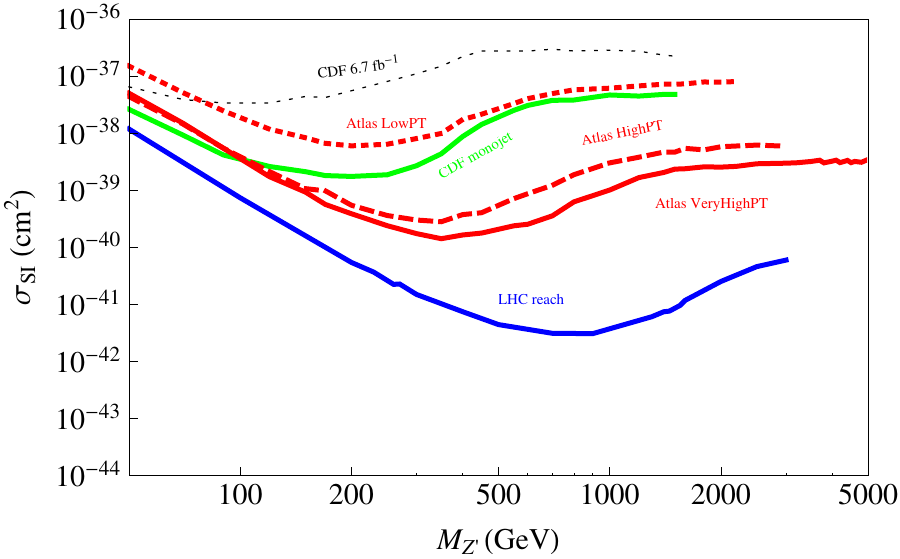}
\caption{Monojet+${\etmiss}$ constraints on direct detection cross sections for $g_{Z'}=g_D$ and $\mchi=5$ GeV. The solid, dashed and dotted red curves are for ATLAS Monojet constraints with VeryHighPT, HighPT and LowPT cuts described in Table~\ref{ATLAS_cuts}. The green solid curve is the monojet constraint from CDF. 	 The dashed green and blue curves are constraints from CDF and ATLAS dijet resonance searches. The solid blue curve is LHC 5$\sigma$ reach assuming a centre-of-mass energy of 14 TeV and a luminosity of 100 fb$^{-1}$.
\label{fig:tev_monojet}}}

\FIGURE[h!]{
\includegraphics[scale=1]{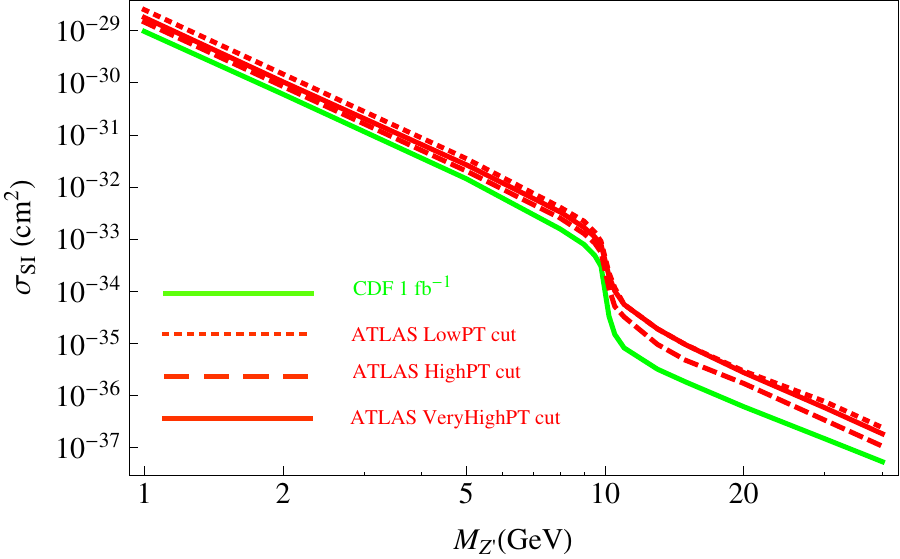}
\caption{Monojet constraints on direct detection cross sections in the case of small $M_{Z'}$, assuming $g_{Z'}=g_D$ and $\mchi=5 $ GeV.
\label{fig:tev_monojet_kink}}}

Assuming $g_{D}=g_{Z'}$, we convert these constraints into limits on the direct detection cross section.
The constraints on the spin independent cross section from Tevatron and LHC monojets searches ~\cite{Aaltonen:2008hh}, for fixed dark matter mass $\mchi=5$ GeV, are shown  in Figs.~\ref{fig:tev_monojet} and \ref{fig:tev_monojet_kink}. Very recently, this search has also been carried out using CDF Run II data set with a luminosity of $6.7$ fb$^{-1}$~\cite{CDF7ifb:001}. A different signal region was chosen, which is also shown in Table.~\ref{ATLAS_cuts}. They did a binned study in the signal region, and they translated their constraints on the generator level rate of the monojet + MET in the signal region for mediator masses of 100 GeV and 10 TeV, respectively. To incorporate it into our study with general mediator masses  other than the two chosen values,  we do a interpolation to get the corresponding constraints. The corresponding constraints on direct detection cross section is shown as the dotted black curve in Fig.~\ref{fig:tev_monojet}, where we can see that the new cuts is different from the one set by ATLAS with VeryHighPT cuts and the previous CDF cuts with 1 fb$^{-1}$.

For very large $\mzp$, we can effectively integrate out the $\zp$. The resulting contact interaction provides a good approximation even at LHC energies. In this limit, both the direct detection and monojet+MET cross sections depend on the same combination $\gzp^2 \gd^2/\mzp^4$, therefore the limits shown in Fig.~\ref{fig:tev_monojet} approach a constant value for very large $\mzp$. We can also see that the contact-interaction limit is reached at larger $\mzp$ for searches at higher energies and more sensitive cuts, as expected. The limits become stronger for intermediate values of $\mzp$, since in this regime, the $\zp$ can be produced on-shell, leading to a significantly enhanced cross section for the monojet+MET process. When $\zp$ mass is comparable or less than the kinematical cuts used in the searches, the monojet+MET cross section starts to be less sensitive to $\mzp$. In this regime, the monojet searches are effectively setting limits on $\gzp^2$, while direct detection still depends on  $\gzp^2 \gd^2/\mzp^4$. Therefore, the limits becomes weaker in this range of $\mzp$, as shown in Fig.~\ref{fig:tev_monojet}. The constraints for very small $\mzp$ is shown in Fig.~\ref{fig:tev_monojet_kink}. We see that in this case, the constraints from collider searches are weak, mainly due to the $\mzp^{-4}$ dependence on the direct detection cross section. As we will see later in this paper, the collider search to $\zp$-like resonances can not provide useful constraint in this regime either.  It remains a challenge to find better probes for such light $\zp$ with only hadronic decay modes. The ``kink" feature in Fig.~\ref{fig:tev_monojet_kink} is due to the threshold effect around the point at which $2 \mchi > \mzp$, where the signal process can only proceed through an off-shell $\zp$.

\FIGURE[h!]{
\includegraphics[scale=1]{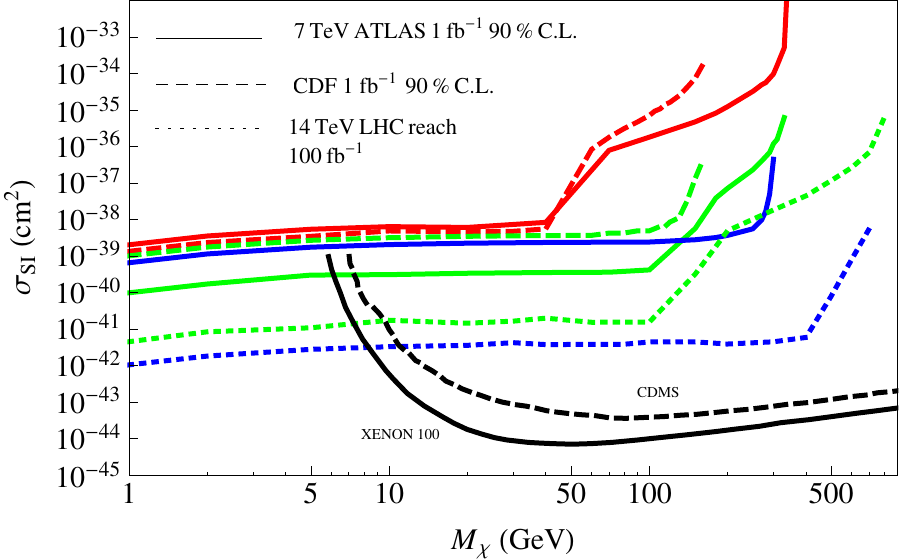}
\caption{Monojet constraints on direct detection cross section. The red, green, blue curves are for $M_{Z'} = $ 100, 300, 1000 GeV, respectively,
together with XENON 100 and CDMS constraints.
\label{fig:tev_monojet_MD}}}

Fig.~\ref{fig:tev_monojet_MD} shows the Tevatron and LHC constraints with several fixed values of $M_{Z'}$, as a function of dark matter mass $M_\chi$.
In the case of $M_\chi$ smaller than around 5 GeV, the constraints from collider physics become stronger than that from the direct detection experiments.
In the regime where $\mchi \ll q$, where $q \sim \mbox{max} (\mzp,p_T^{\rm cut })$  is the hardest scale in the process, the production rate is approximately independent of $\mchi$, and is a function of $\gzp$, $\gd$, and $\mzp$. Given a set of $\gzp$, $\gd$, and $\mzp$, direct detection can depend further on $\mchi$ through the dark matter nucleon reduced mass $M_{*} = M_{N} \mchi / (M_{N} + \mchi )$. However, this dependence is rather weak for $\mchi \sim \mathcal{O} (10)$ GeV since $M_{*} \sim M_N$. Taking together, we expect the limits derived from collider searches are rather insensitive to the dark matter mass $\mchi$. In contrast with the steep weakening of the direct detection bound for light dark matter, collider searches are particularly powerful in this regime.
For heavier dark matter, the visible "kink"-like feature around $2 \mchi \simeq \mzp$ in the curves are due to the transition from $2 \to 2 $ production process $p p (\bar{p}) \to \zp + $jet followed by decay $\zp \to \chi \chi$, to $2 \to 3$ production process $p p (\bar{p}) \to (Z^{\prime *} \to \chi \chi)  + $jet which has a much smaller production rate. For example, there is such a feature on the red curve in Fig.~\ref{fig:tev_monojet_MD} near $\mchi \sim 60 $ GeV for $\mzp = 100 $ GeV. One can see that this turning point also shows up at the green and blue dotted curves. Notice there is also a second turning point, for example $\sim 300$ GeV for the 7 TeV LHC,   at which the dark matter is too heavy to be produced. 

The  new physics search potential in the $pp\rightarrow$ monojet + MET channel has been studied  at 14 TeV center-of-mass energy with luminosity of 100 fb$^{-1}$ in Ref.~\cite{Vacavant:2001sd}. With the requirement that MET is larger than 500 GeV, they found that the SM background is about $B=2\times10^{4}$ events.
For a 5$~\sigma$ discovery, we require that the signal should be larger than than $5\sqrt{B}$. The  5~$\sigma$ reach is shown in Fig.~\ref{fig:tev_monojet} for a light WIMP and in Fig.~\ref{fig:tev_monojet_MD} for $M_{Z'}$ fixed to be 300 and 1000 GeV.
 One can see that for $M_{Z'}$ around $300\sim1000$ GeV, LHC can potentially reach the interesting region where anomalies are claimed by CoGeNT~\cite{Aalseth:2011wp,Aalseth:2010vx} and CRESST~\cite{Angloher:2011uu} direct detection.

\subsection{Constraint for the spin dependent case}
\label{sec:sd}

\FIGURE[h!]{
\includegraphics[scale=1]{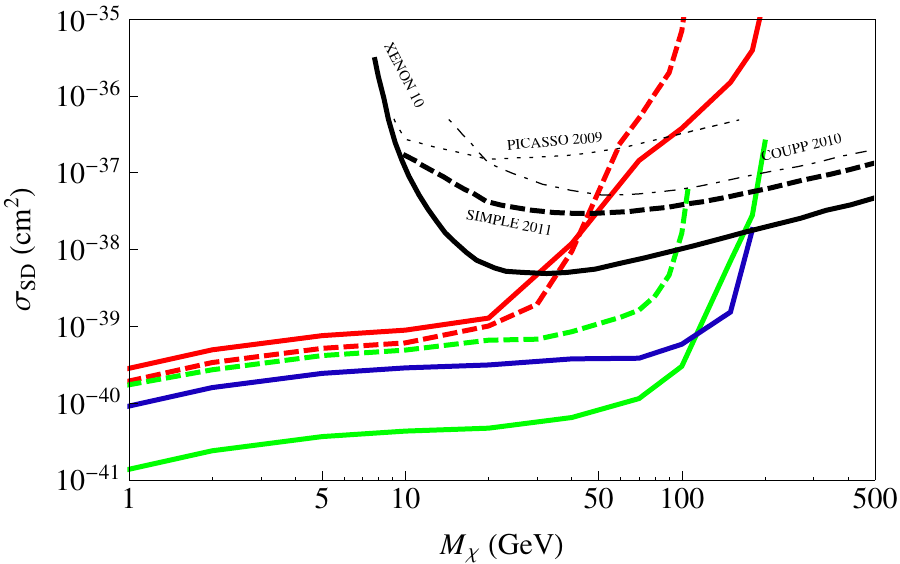}
\caption{ Constraints on spin-dependent direct detection cross section between dark matter particle and the nucleon (proton or neutron). The solid and dashed
coloured curves are for ATLAS monojet constraint with VeryHighPT cut and CDF monojet constraint, respectively. The red, green, and blue curves are for $M_{Z'}=100,300,1000$ GeV respectively. The black solid curve is the constraint for neutron scattering from Xenon10~\cite{Angle:2008we}, the thick dashed curves is that of  proton scattering
from SIMP2010~\cite{Felizardo:2010mi}. The thin dotted dash curves and dotted curves are from COUPP 2010~\cite{Petriello:2008jj} and
PICASSO 2009~\cite{Archambault:2009sm} on the constraint for proton scattering. The constraint on $\sigma_{\rm SD}$ for the
neutron can be got from the constraining curves scaled by a factor of $(\Delta\Sigma_{n}/\Delta\Sigma_{p})^2$, which are defined in the caption of Table~\ref{tab:zprime_int}.
\label{fig:SD}}}

We have so far concentrated on the cases that $Z'$ couples only to the vector currents of quarks
and dark matter. If $Z'$ couples to the axial vector current of dark matter particle,
the direct detection cross section depends on either its momentum or the nucleon axial charge $\Delta \Sigma$.
If dark matter particle is a Majorana fermion, it can only couple through an axial vector current
and only operators of $O_2$ amd $O_4$ in Table~\ref{tab:zprime_int} are relevant for the direct detection.
We generically expect both of these operators to be present. If $Z'$ couples only to the vector current of quark fields ($O_2$),
scattering amplitude is proportional to $|\vec p|$, and the direct detection cross section is suppressed by a factor of $v^2\sim 10^{-6}$.
Therefore, we focus on the case in which $O_{4}$ dominates. The collider signal is largely insensitive to the
details of the couplings since the typical momentum exchange there is the hard scale of scattering process.
 Assuming $g_{Z'}=g_{D}$, the collider constraint on the spin-dependent is shown in Fig.~\ref{fig:SD}.
 For the dark matter mass smaller than around 100 GeV, Tevatron constraint is already much stronger than that from direct detection.
 The dependence on the mass of mediator is similar as in the case of spin-independent scattering.
 The features of the curves are similar to the spin-independent case studied in the previous subsection.

\section{Constraint from Dijet Searches for $\zp$. }
\label{Sec:dijet}

If $Z'$ is sufficiently light and its coupling to quarks is unsuppressed,  direct  searches for its collider signal, such as a resonance in the di-jet final state, should be promising. There are two main approaches to search for $Z'$ in the di-jet channel. First, one can perform a ``bump hunting"
in the dijet invariant mass distribution.  One can also look for deviations in other jet kinematical distributions,
such as angular correlation. In this section, we consider the collider limits on both searches and their constraint
on the direct dark-matter detect cross section.

\subsection{Dijet resonance search with $g_{Z'}=g_D$}

CDF collaboration has studied the dijet final state constraints for $Z'$ model with 1.13 fb$^{-1}$ data~\cite{Aaltonen:2008dn}. ATLAS collaboration has also published a study of dijet search of new resonances with a luminosity of 1 fb$^{-1}$~\cite{Aad:2011fq}. In both studies, they fit the dijet invariant mass spectrum with the function $f(x)=p_{1}(1-x)^{p_{2}}x^{p_{3}+p_{4}\ln x}$ to model the QCD background, where $p_i$ are parameters and $x=m_{jj}/\sqrt{s}$ with $m_{jj}$ the dijet invariant mass and $\sqrt{s}$ the center-of-mass energy of the collider. In our study, we use the Bayesian method described in Ref.~\cite{Choudalakis:2011bf}. We first assume that the prior probability density function of the coupling $g_{Z'}$ to be a uniform function between 0 and $4\pi$. For each value of $g_{Z'}$,  we simulate the signal and calculate the likelihood function, from which the posterior probability density function is obtained.

The constraints on the direct detection cross section from CDF and ATLAS, assuming $g_{Z'} = g_D$ and $\mchi=5$ GeV, are shown as the green and blue solid curves in Fig.~\ref{fig:tev_dijet}. In the region $900 {~\rm GeV} < M_{Z'} < 4$ TeV, dijet resonance search in ATLAS gives the most stringent constraint, while for $M_{Z'}$ around 300 GeV to 900 GeV, the CDF dijet resonance search gives the most stringent constraint.

\FIGURE[th]{
\includegraphics[scale=1]{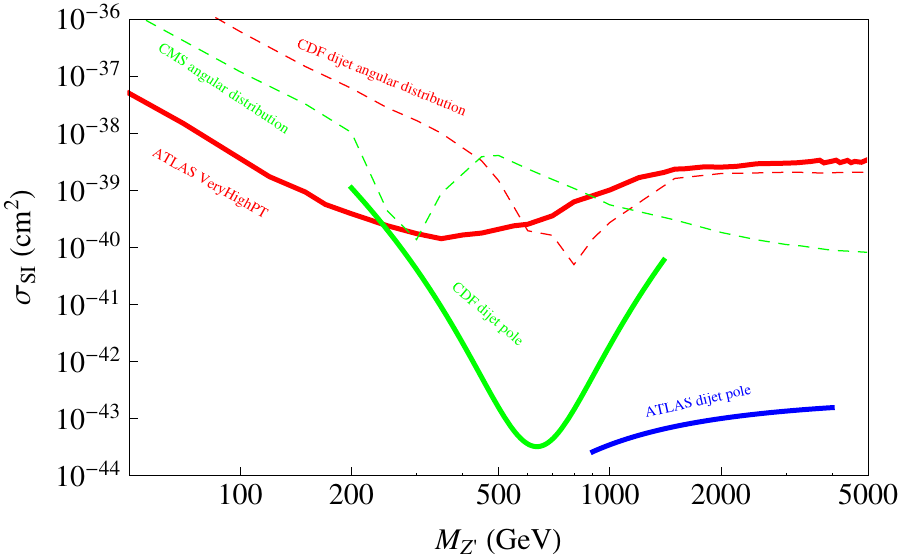}
\caption{Constraints on spin-independent direct detection cross sections assuming $g_{Z'}=g_D$ and $\mchi =5$ GeV. The solid red curve is the constraint from ATLAS monojet constraints with VeryHighPT cuts described in Table~\ref{ATLAS_cuts}. The green solid curve is the bound from di-jet resonance search by CDF. 	The dashed green and red  curves are constraints from CMS and D0  dijet angular distribution measurements. The solid blue curve is the constraint from ATLAS di-jet resonance search with 1 fb$^{-1}$ data set.
\label{fig:tev_dijet}}}

The corresponding constraints in the case of heavier dark matter can also  be obtained by simple scaling of the constraints presented in Fig.~\ref{fig:tev_dijet}. In the case of $g_D = g_{Z'}$,  the constraint from dijet resonance searches depends very weakly on the mass of DM since the $\zp \to \chi \chi$ decay channel only contributes $\sim 5 \% $ of the width of $Z'$. Therefore, the threshold effect  on $\Gamma_{\zp}$ at $\mzp \sim 2 \mchi$   is negligible, and the constraints on $\zp$ mass and couplings from CDF and ATLAS dijet resonant searches are also applicable to the case of heavy DM. Therefore,
for $\mchi \neq 5$ GeV, one could just scale the di-jet  constraints from Fig.~\ref{fig:tev_dijet} by a factor of $\frac{ \mchi^{2}(M_{N}+5)^{2}}{5^{2}(M_{N}+\mchi)^{2} }$. 

As shown in Fig.~\ref{fig:tev_dijet}, In the region between $200 $ GeV $< \mzp < 4 $ TeV, the dijet resonance searches provide significantly stronger constraints  than the mono-jet + MET searches.

\subsection{$g_{Z'}\neq g_{D}$}

In some sense, the assumption $g_D = g_{Z'}$ is overly simplistic
in the dijet case, because at large $M_{Z'}$, the dijet cross section
constraints only $g_{Z'}$ with little effect on $g_D$. In fact,
the dijet constraint is so strong that $g_D$ can have a very large
value before runs into contraction with monojet constraint. A large
$g_D$ certainly weakens the direct detection constraint considerably.

Let us study the effect of allowing $\gzp \neq \gd$. The constraints for $g_{D}/g_{Z'}=1,3,5,10,20$ for monojet and dijet resonance studies are shown in Fig.~\ref{fig:interplay}. The dependence of dijet cross section on $g_D$ is only through next leading order effects, such as the $\zp$ width. Therefore, we expect the dijet constraint on $g_{Z'}$ depends on $g_D$ weakly unless the invisible width of $Z'$ dominates over the visible in which case the constraint on $g_{Z'}$ becomes weaker. Therefore, we can see that the constraint on direct detection cross section still scales roughly as $g_D^2$ (or slightly stronger), and it becomes much weaker in the case of $g_D \gg g_{Z'}$. On the other hand, the monojet cross section is proportional to $g_{Z'}^2 g_D^2$ unless decay channel $\zp \to \chi \chi$ dominates. Therefore, one expects that the constraint on direct detection cross section does not change much with $g_D$. From Fig.~\ref{fig:interplay}, we see that in the region $g_{D}\gg g_{Z'}$, monojet constraints may dominate over dijet constraints.

\FIGURE[th]{
\includegraphics[scale=1]{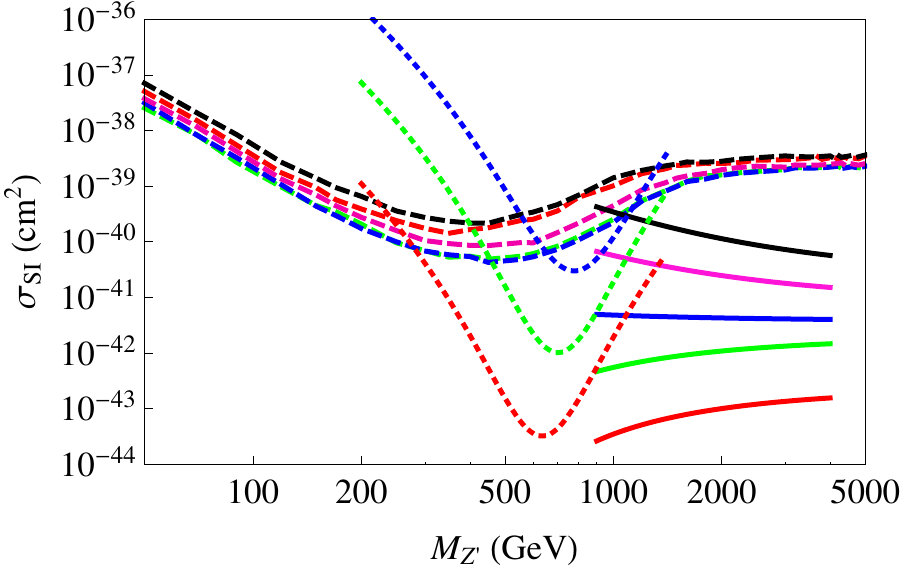}
\caption{Comparing monojet and dijet constraints. The solid, dashed and dotted curves are for ATLAS dijet resonance search, ATLAS monojet search with VeryHighPT cut and CDF dijet search, respectively. The red, green, blue, pink and black are for $g_D/g_{Z'} = 1,3,5,10,20$, respectively. The mass of DM is assumed to be 5 GeV.
\label{fig:interplay}}}

\subsection{Dijet Angular Distribution}

CDF, D0, ATLAS and CMS collaborations have done searches for quark compositeness through study of dijet angular distribution~\cite{Abe:1996mj,Aba:2009mh,Aad:2011aj,Khachatryan:2011as}. The quantities of interest is the normalized differential cross section $\bar\sigma = (1/\sigma_{\rm dijet})(d \sigma_{\rm dijet} / d\chi_{\rm dijet})$, where $\chi_{\rm dijet} = (1+|\cos\theta^*|)/(1-|\cos\theta^*|)$ and $\theta^*$ is a scattering angle in the rest frame of the two partons.

In our study, the QCD background is simulated using Pythia 8.1.4.5~\cite{Sjostrand:2007gs}, and Fastjet ~\cite{Cacciari:2006sm}. 
We use AntiKT jet algorithm with $R = 0.4$. The new physics contribution is simulated using a private code~\cite{Hanlib}.
 For each group of dijet centre-of-mass energy defined in Refs.~\cite{Aba:2009mh,Khachatryan:2011as}, we calculate the $\chi^{2}$ which defined as
\begin{equation}
\chi^{2} = \sum_{i} \frac{(\bar\sigma^{\rm new}_{i} + \bar \sigma^{\rm QCD}_{i} - \bar\sigma^{\rm exp}_{i})^{2}}{\delta_{\rm exp}^{2}+\delta_{\rm QCD}^{2}}\ ,
\end{equation}
where $\bar\sigma^{\rm new}_{i}$, $\bar\sigma^{\rm QCD}_{i}$ and $\bar\sigma^{\rm exp}_{i}$ are the new contributions, QCD background and experimental value in the $i$-th bin for certain $M_{jj}$ group, respectively. $\delta_{\rm exp}$ and $\delta_{\rm QCD}$ are the uncertainties of experimental values and QCD background. To get 95\% C.L. constraint on $g_{Z'}$ for certain values of $g_{D}$ and $M_{Z'}$, we require that in each $m_{jj}$ group the possibility to get calculated $\chi^{2}$ should be smaller than 0.05.
 The constraints on $g_{Z'}$ from CMS and D0 are shown in Fig.~\ref{fig:dijet_angular}, where the red and green curves are for D0 and CMS respectively; and the corresponding constraints on direct detection cross sections are shown in Fig.~\ref{fig:tev_dijet}. 

\FIGURE[h!]{
\includegraphics[scale=1]{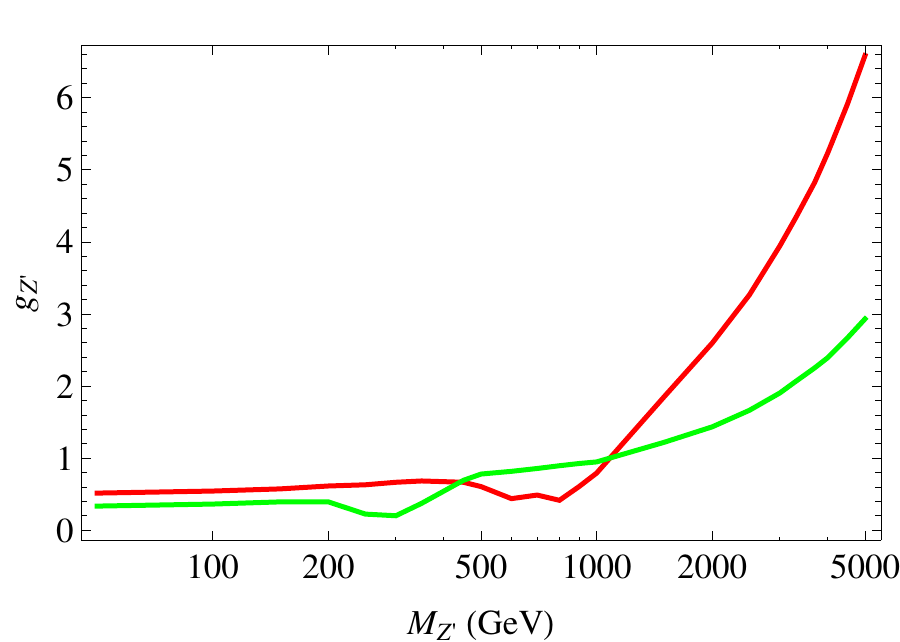}  \\
\caption{ The constraints on the $\zp$ coupling, $\gzp$, from di-jet angular distribution measurements at D0 (red) and CMS (green).
\label{fig:dijet_angular}}}

\FIGURE[h!]{
\includegraphics[scale=1]{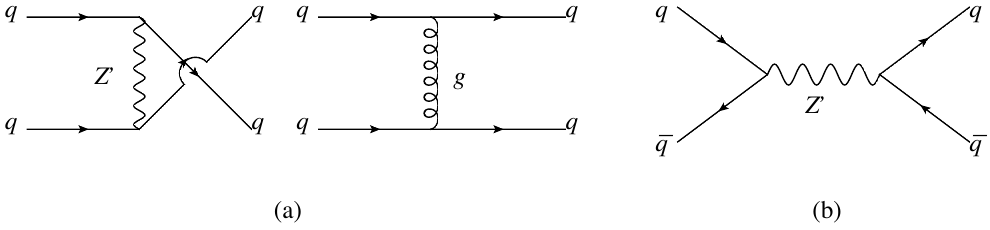}
\caption{Dominant contributions to dijet angular distribution at LHC.
\label{fig:dijet_diagram}}}

Since Tevatron is a $p\bar p$ collider, the main background is from $q\bar q \rightarrow  j j$ and $g g\rightarrow j j $. The dominant contribution to the signal is from $q\bar q \rightarrow Z' \rightarrow q \bar q$, where $Z'$ can be either on or off shell.  $gg\rightarrow gg$ provides dominant background in the energy region of $\sqrt{\hat s} < 300$ GeV. However, it drops steeply at $\sqrt{\hat s} \simeq 500$ GeV, where $q\bar q \rightarrow jj$ becomes dominant with a much smaller rate.  At the same time,  $\zp$ with $\mzp \sim 500 $ GeV can still be produced on-shell. Therefore, we see from red curve in Fig.~\ref{fig:dijet_angular} that the constraint gets stronger at around 500 to 800 GeV.
For larger $M_{Z'}$, $\zp$ on-shell production is strongly suppressed by the steeply falling PDF. As a result, the constraint on the coupling gets weaker and eventually reaches the limit of the contact interaction, which is illustrated by the plateau of the red dashed curve in Fig.~\ref{fig:tev_dijet}. The height of the plateau can be interpreted as $\Lambda\approx 2$ TeV for a quark composite operator $(2\pi / \Lambda^2 ) (\bar q \gamma_\mu q)^2$ which agrees with the result from the compositeness search at D0~\cite{Aba:2009mh}.

At the LHC, the major background comes from $gg\rightarrow jj$ and $qq\rightarrow jj$. The signal contains two contributions which are shown in Fig.~\ref{fig:dijet_diagram}, where (a) is an $1/N_C$ suppressed interference between QCD and the new physics which is enhanced by the parton distribution function of the valence quarks; and (b) is suppressed by parton distribution function of $\bar{q}$ but will still be significant if $Z'$ can be produced on shell. It turns out that in the case of light $Z'$ with a mass smaller than around 500 GeV,  the contribution from $q\bar q$ initial state dominates. Since the first energy bin for this study in CMS starts from $m^{jj}_{\rm min}=250$ GeV, therefore we expect the resonance enhanced  $q\bar q$ contribution to the angular distribution can only be significant when $\mzp$ is comparable with $m^{jj}_{\rm min}$.  This is the reason for the local enhancement of the constraint around $\mzp \sim 250$ GeV, shown in the green curve in Fig.~\ref{fig:dijet_angular}. There is no such a dent in the red curve since for the D0 study, the constraint is always from the last few bins of the dijet center of mass energy.

The constraint on $\gzp$ from dijet angular distributions is not as sensitive to the width of the mediator as in the case of dijet resonance search. Therefore, the constraint on direct detection cross section can be estimated by simply scaling the curves in Fig.~\ref{fig:dijet_angular} by a factor of $g_{D}^{2}/g_{Z'}^{2}$.

From Fig.~\ref{fig:tev_dijet}, we can see that the current constraints from dijet angular distribution search is not as strong as the one from dijet resonant search.
In longer term, we expect angular correlation measurement will be more relevant for the case of very large $\mzp$.

\section{Conclusion}
\label{Sec:conclusion}

In this work, we consider the possibility of probing light dark matter at the colliders.  We interpret the current limits and potential search reach at the colliders as constraints on the direct detection cross section. As already been demonstrated in earlier works, such an approach can yield stronger, albeit model dependent, constraint in comparison with the direct detection for the light dark matter $M_\chi \sim 5$ GeV. The simplest approach to connect the collider and direct detection limits is to work in the limit in which the mediator is so heavy that it can be integrated out both for direct detection and for collider searches at the Tevatron and the LHC. Focusing the monojet+MET observable, the connection between these two set of observables can be established in a straightforward manner.

We argue that while using contact operators is probably the most straightforward approach with the most direct connection between dark matter direct detection and collider searches, it is important to consider the case in which the mediator of dark matter and SM interactions will also be accessible at high energy colliders. Earlier studies have emphasized the effect of light mediator on the monojet + MET channel.  In this work,  we adopt the strategy of combining the search in direct dark matter production channels of mono-jet(photon) + MET, and the search of mediator directly at the colliders. We demonstrate that the two approaches are complementary.

As a concrete example, we study the case in which the dark matter is a Dirac fermion with its interaction with the SM mediated by a spin-1 $Z'$ through vector like couplings. Such a $Z'$ can be searched directly in dijet channels. We studied the constraints on the $\zp$ model from both dijet  and mono-jet + MET searches at the Tevatron and the LHC. In the case of a strongly coupled $\zp$, we have carefully taken into the effect of the finite resonance width. We then combined the constraints in dijet resonance searches with the mono-jet + MET searches, and interpreted them as constraints on the direct detection cross sections. In the case of spin-independent interaction, the monojet constraint is stronger than the direct detection constraints only in the case that $M_\chi$ is smaller than about $5$ GeV. At the same time, if the mass of the mediator is larger than around 250 GeV, the constraint from dijet search is stronger than the monojet constraints. In the case of spin-dependent interaction, the collider constraints become much stronger then the direct detection constraints.

The process of mono-photon + MET is also studied in CDF~\cite{Aaltonen:2008hh} and a theoretical study has been done for LHC in Ref.~\cite{Fox:2011pm}, which shows that the mono-photon constraints are relatively weaker than the mono-jet constraints due to that the rate is much smaller although the mono-photon signal is cleaner than monojet.

\section{Acknowledgement}

We acknowledge A. Pukhov and K. Zurek for illuminating discussions. H.A. would like to thank X.-G. He and M. Ramsey-Musolf for useful discussions. H.A. and L.T.W would like to thank Institute of Nuclear and Particle Physics at Shanghai JiaoTong University for their hospitality,  part of the work presented here has been carried out during their visits to the institute. H. A. and X. J. 's research at University of Maryland are supported by the U. S. Department of Energy via grant DE-FG02-93ER-40762. H.A.'s research at Perimeter Institute is supported in part by the Government of Canada through NSERC and by the Province of Ontario through MEDT. X. J. is supported in part by Shanghai lab for particle physics and cosmology and a 985 grant from Shanghai Jiao Tong university. L.T.W. is supported by the NSF under grant PHY-0756966 and the DOE Early Career Award under grant DE-SC0003930.

\appendix
\section{Propagator with large width}
\label{Appendix:A}

In the prototype model described in this work, $Z'$ is assumed to couple all the quarks universally. Therefore, if $g_{Z'}$ becomes large, $Z'$ becomes a broad resonance and the Breight-Wigner approximation can no longer be used as a good estimation. In the broad resonance case, the kinetic dependence should be included in the propagator, which can be written as
\begin{equation}\label{Eq:propagator}
\frac{i}{s-m^{2}+i\sqrt{s}\Gamma_{\rm tot}(s)}\ ,
\end{equation}
where $i\sqrt{s}\Gamma_{\rm tot}(s)$ includes all imaginary parts in the two-point correlation function of $Z'$. The diagrams are shown in Fig.~\ref{fig:bubbles}. The diagrams can be separated into two groups, in the first group shown in the first line of Fig.~\ref{fig:bubbles}, it is easy to see that the each diagram is proportional to $(g^{2}N)^{r}$, where $g$ is the coupling between the intermediate particle and $Z'$, $N$ is the number of fermion degrees of freedom and $r$ is the number of fermion loops. In our case, since the $Z'$ universally couples to all quarks, there are at least 15 light degrees of fermions. Therefore, one can see that the width of $Z'$ becomes broad when $g$ is order one. And the summation of the first line in Fig.~\ref{fig:bubbles} dominates over others. After summing up all the leading order diagrams, the propagator can be written as in Eq.~(\ref{Eq:propagator}). In the case that $\sqrt{s} \gg m_{f}$ where $m_{f}$ is the mass of the fermion in the loop, $\Gamma_{Z'}(s)$ is proportional to $\sqrt{s}$ and can be estimated as
\begin{equation}
\Gamma_{\rm tot}(s) \approx \frac{\sqrt{s}}{M_{Z'}} \Gamma_{Z'}\ ,
\end{equation}
where $\Gamma_{Z'}$ is the physical width of $Z'$.

Fixed width is used in most of packages of simulation which might not be very proper in dealing with broad resonances. A comparison between propagators with kinetic width and fixed width are shown in Fig.~\ref{fig:propagator}, where the solid and dashed curves are for the kinetic and fixed widths cases, respectively. Different colors are for different choices of the 't Hooft coupling defined as $g\sqrt{N}$ changing from 2 to 10. One can see that in the case of $g\sqrt{N}=10$, the deviation of the fixed width one from the kinetic one can be as large as a factor of ten. In our simulation, we modified the CalcHEP~2.5.7 under the instruction the authors \footnote{We thank A. Pukhov for help in modifying CalcHEP. In particular, the relevant modifications are made in the file sqme\_aux.inc in the folder c\_source/sqme\_aux to invoke the kinetic width. } so that the effect of kinetic width can be realized.

\FIGURE[th]{
\includegraphics[scale=1]{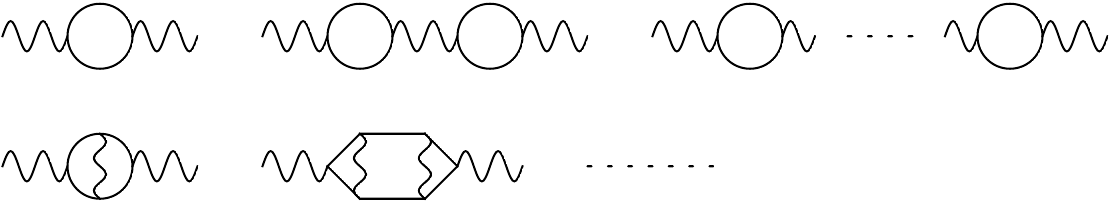}
\caption{Diagrams of $Z'$ two-point correlation function.
\label{fig:bubbles}}}

\FIGURE[th]{
\includegraphics[scale=1]{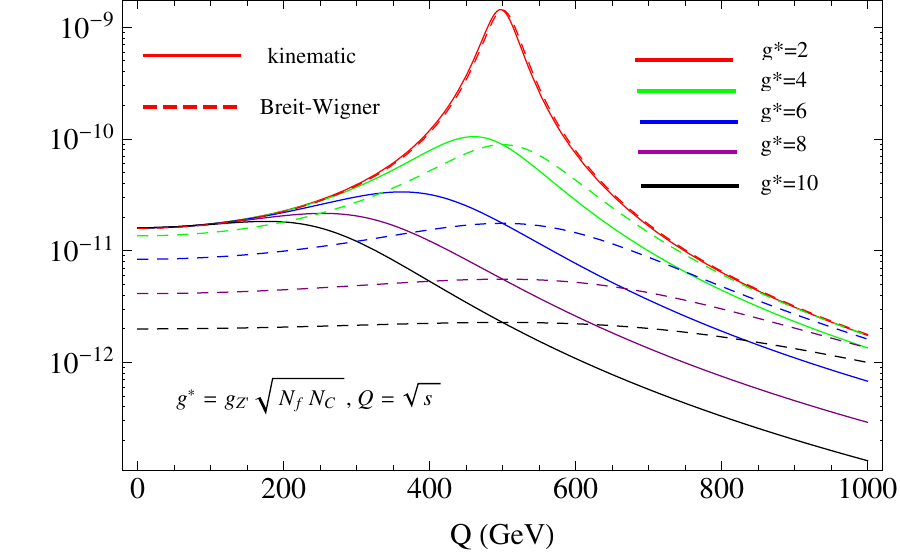}
\caption{Comparison between the kinetic propagator (solid curves) and the fixed width ones (dashed). Red, green, blue, purple and black curves are for $g^* = 2,4,6,8,10$, respectively. The curves are showing the absolute square of the propagator which is $\frac{1}{(q^{2}-M_{Z'}^{2})^{2}+\left(\frac{g^{2}N}{12\pi}\right)^{2}s\Gamma^{2}(s)}$.
\label{fig:propagator}}}

\section{Constraints on couplings of $Z'$}
\label{Appendix:B}

\FIGURE[th]{
\includegraphics[scale=1]{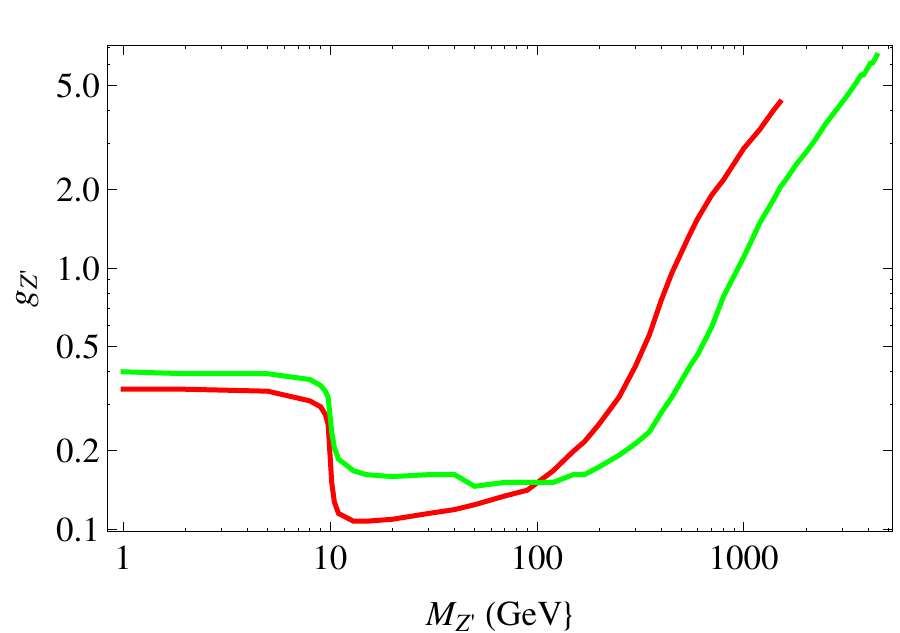}
\caption{Constraints on $g_{Z'}$ from monojet studies at CDF (red curve) and ATLAS (green curve), assuming $M_\chi=5$ GeV and $g_{Z'}=g_D$.
\label{fig:gzp1}}}

\FIGURE[th]{
\includegraphics[scale=1]{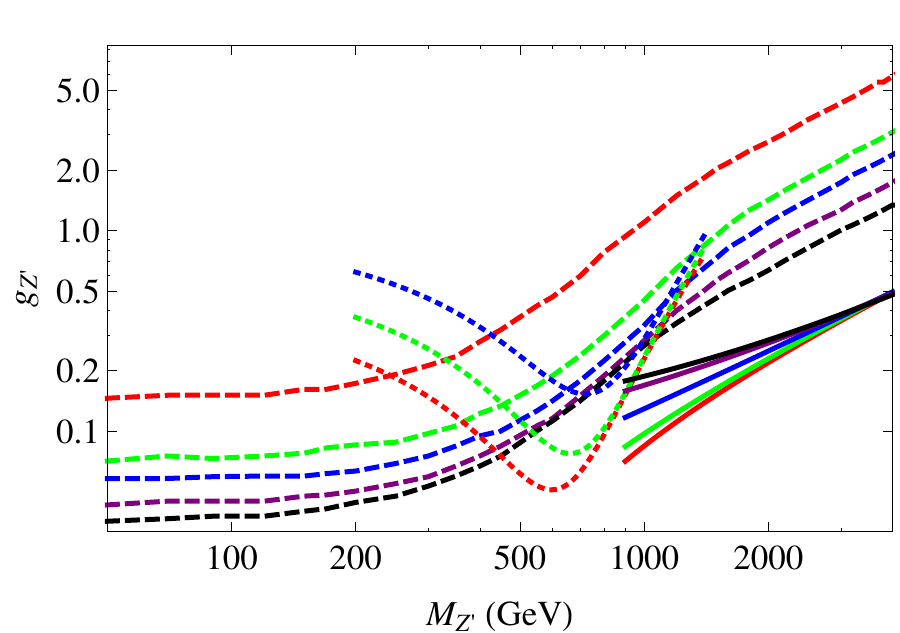}
\caption{Monojet and dijet constraints on $g_{Z'}$ for $g_{D}/g_{Z'} = 1,3,5,10,20$, which are labeled by red, green, blue, purple, black curves respectively. The solid, dotted and dashed curves are for ATLAS dijet resonance search, CDF dijet resonance search and ATLAS monojet search with VeryHighPT cut, respectively. We fix $M_\chi=5$ GeV.
\label{fig:gzp2p}}}

We present the constraints on $g_{Z'}$ from monojet and dijet experiments. In Fig.~\ref{fig:gzp1}, we show the red and green curves
corresponding to monojet-only constraints from CDF and ATLAS with VeryHighPT cut, respectively, asssuming $M_\chi=5$ GeV and $g_{Z'}=g_D$.
We see that for $M_{z'}$ smaller than $2M_\chi$, the constraint is around $0.4$. For $M_{Z'}$ between $2M_\chi$ and 100 GeV,
because of the resonance production, the constraint is much stronger. For $M_{Z'}>200$ GeV, the constraint
is in the effective theory form $g_{Z'}g_D/M_{Z'}^2$, or $g_{Z'}$ is linear in $M_{Z'}$. Apparently, ATLAS constraint
is better in the high-$M_{Z'}$ region. For the same low $M_\chi$, the dependence on the ratio of $g_{D}/g_{Z'}$
is shown in Fig.~\ref{fig:gzp2p} where the solid and dotted curves are now constraints from dijet studies of ATLAS and CDF,
whereas the dashed curves show the constraints from ATLAS monojet study with VeryHighPT cut. The red, green, blue, purple and black colors
are corresponding to $g_{D}/g_{Z'}=1,3,5,10,20$, respectively. Two important features can be commented on. First,
at large $M_{Z'}$, the dijet contraint is much stronger than monojet, particularly from LHC. At very large $M_{Z'}$, the
dijet constraint is mostly on $g_{Z'}$ independent of $g_{D}$. As $M_{Z'}$ becomes smaller, $g_D$ affects the width
of $M_{Z'}$, larger $g_{D}$ weakens the constraint on $g_{Z'}$. However, to weaken this constraint
to the same level as the monojet, one would allow an extremely large $g_{D}$, beyond the effective range of the
$Z'$ model. Thus the tightest constraint happens for $g_{Z'}$ from dijet and $g_{D}$ from a reasonable theoretical bound.
Second, at $M_{Z'}$ less than few hundred GeV, the constraint is mostly from monojet, and the relevant parameter
is just the product $g_{D}g_{Z'}$. All results are nearly independent of dark matter mass so long as $M_\chi<50$ GeV or so.

\bibliographystyle{jhep}
\bibliography{lcdm}

%

%
\end{document}